\documentclass[aps,twocolumn,superscriptaddress,preprintnumbers]{revtex4}

\usepackage{xr-hyper}
\usepackage{mathrsfs, hyperref}
\usepackage{amssymb, amsbsy, amsmath, latexsym, dsfont, array, layout, graphics,mathrsfs,braket,amsfonts,amsthm,
  amssymb,graphicx,subfigure, youngtab,color,bm,mathtools,braket,verbatim,url,cleveref,natbib,hypernat,extarrows,inputenc,multirow,bbm}

\usepackage[usernames,dvipsnames]{xcolor}
\usepackage{graphicx,psfrag}
\usepackage{amsfonts}
\usepackage[figuresright]{rotating}  
\usepackage{amssymb}
\usepackage{amsmath}
\usepackage{subfigure}
\usepackage{multirow}
\usepackage{tabularx}
\usepackage[resetlabels]{multibib}
\newcites{supp}{Supplementary References}
\usepackage{amssymb, amsbsy, amsmath, latexsym, dsfont, array, layout, graphics,mathrsfs,braket,amsfonts,amsthm,
  amssymb,graphicx,subfigure,dcolumn, youngtab,color,bm,mathtools,braket,verbatim,url,cleveref,natbib,hypernat,extarrows,inputenc,multirow}
\usepackage[usernames,dvipsnames]{xcolor}

\newcommand{\sg}[1]{{\color{black} #1}} 
\usepackage[mathlines]{lineno}

\begin{document}
\title{Dynamics of self-dual kagome metamaterials and the emergence of fragile topology}

\author{Pegah Azizi}
\affiliation{%
Department of Civil, Environmental, and Geo- Engineering,
University of Minnesota, Minneapolis, MN 55455, US}
\author{Siddhartha Sarkar}
\affiliation{%
Department of Physics, University of Michigan, Ann Arbor, MI 48109, USA
}
\author{Kai Sun}
\affiliation{%
Department of Physics, University of Michigan, Ann Arbor, MI 48109, USA
}
\author{ Stefano Gonella}%
\email{sgonella@umn.edu}
\affiliation{%
Department of Civil, Environmental, and Geo- Engineering,
University of Minnesota, Minneapolis, MN 55455, US}%
\date{Accepted 22 March 2023; published 12 April 2023; updated on arXiv 28 January 2025}
\text{Published DOI:\href{https://journals.aps.org/prl/abstract/10.1103/PhysRevLett.130.156101}{10.1103/PhysRevLett.130.156101}}
\begin{abstract}
Recent years have seen the discovery of systems featuring fragile topological states. These states of matter lack certain protection attributes typically associated with topology and 
are therefore characterized by weaker signatures that make them elusive to observe. Moreover, they are typically confined to special symmetry classes and, in general, rarely studied in the context of phononic media. In this Letter, we theoretically predict the emergence of fragile topological bands in the spectrum of a twisted kagome elastic lattice with three-fold rotational symmetry, in the so-called self-dual configuration. A necessary requirement is that the lattice is a structural metamaterial, in which the role of the hinges is played by elastic finite-thickness ligaments. The interplay between the edge modes appearing in the bandgaps bounding the fragile topological states is also responsible for the emergence of corner modes at selected corners of a finite hexagonal domain, which qualifies the lattice as a second-order topological insulator. We demonstrate our findings through a series of experiments via 3D Scanning Laser Doppler Vibrometry conducted on a physical prototype. The selected configuration stands out for its remarkable geometric simplicity and ease of physical implementation in the panorama of dynamical systems exhibiting fragile topology.

\end{abstract}

\maketitle
Topological Insulators (TIs), materials with a gapped band structure characterized by topological invariants, have gained increasing attention due to their unique topologically protected edge dynamics. As a result of the so-called bulk-boundary correspondence, they can support edge and interface modes that are immune to back scattering and robust against perturbations and defects.
Recent studies have introduced a new class of so-called fragile topological states~\cite{po2018fragile,bradlyn2019disconnected,ahn2019failure,bouhon2019wilson,AxionInsulator,song2020twisted,peri2020experimental,wieder2020strong,lian2020landau,bouhon2020geometric,de2019engineering,peri2021fragile}. 
A set of $m$ frequency bands of a Hamiltonian with some symmetries is called fragile topological if the bands cannot be represented by Symmetric (constrained by the symmetries of the Hamiltonian) exponentially Localized Wannier Functions (SLWF) -generalization of atomic orbitals- in real space; but addition of $n$ trivial bands to those bands allows for representation of the total $(m+n)$ bands by SLWF~\cite{po2018fragile,bradlyn2019disconnected,ahn2019failure,song2020twisted}. This is in contrast with conventional topological bands where addition of trivial bands does not trivialize the topology. This fragility of the topology means that there cannot exist robust edge states protected by the nontrivial topology at the boundary. 
However, fragile topological systems in 2D are predicted to have corner modes~\cite{AxionInsulator,ahn2019failure,wieder2020strong,manes2020fragile}. 

Another class of systems of interest in topological mechanics is Maxwell lattices, which have an equal number of degrees of freedom and constraints in the bulk~\cite{doi:10.1080/14786446408643668,CALLADINE1978161} and are therefore on the verge of mechanical instability~\cite{PhysRevLett.103.205503,doi:10.1146/annurev-conmatphys-033117-054235,PhysRevE.83.011111,Lubensky_2015}. 
A typical example 
in 2D is the kagome lattice, whose unit cell consists of two triangles pinned at a vertex and relatively rotated~\cite{doi:10.1073/pnas.1119941109}. 
Several studies have addressed the mechanical properties~\cite{FLECK2007562,Hyun2002,ARABNEJAD2013249,ZHANG20083751,10.1115/1.2913044,doi:10.1073/pnas.1119941109,PhysRevLett.129.204302} and wave propagation characteristics~\cite{PhysRevB.98.094302,doi:10.1063/1.5045837,doi:10.1063/1.4921358,doi:10.1121/1.2179748} of these lattices under a variety of cell shapes and effective hinge conditions. Relevant work has studied the faith of mechanical and topological properties in the transition from ideal configurations featuring perfect hinges to structural lattices, as they would be obtained via machining, cutting or 3D printing. It has been shown that the topological polarization is preserved, albeit diluted, and the zero modes are shifted to finite frequencies~\cite{PhysRevApplied.16.064011,PhysRevLett.121.094301}. Fruchart et al. \cite{Fruchart2020} showed that twisted kagome lattices exhibit a special type of duality, whereby a hidden symmetry guarantees that any pair of configurations that are symmetrically located (in configuration space) with respect to a critical configuration referred to as \textit{self-dual}, display identical phonon spectra - a condition that is however relaxed working with non-ideal lattices of beams \cite{PhysRevB.102.140301}. The self-dual case presents peculiar dynamics, with a two-fold degenerate spectrum over the entire Brillouin Zone (BZ). 

\begin{figure}[t!]
\includegraphics[width=\columnwidth]{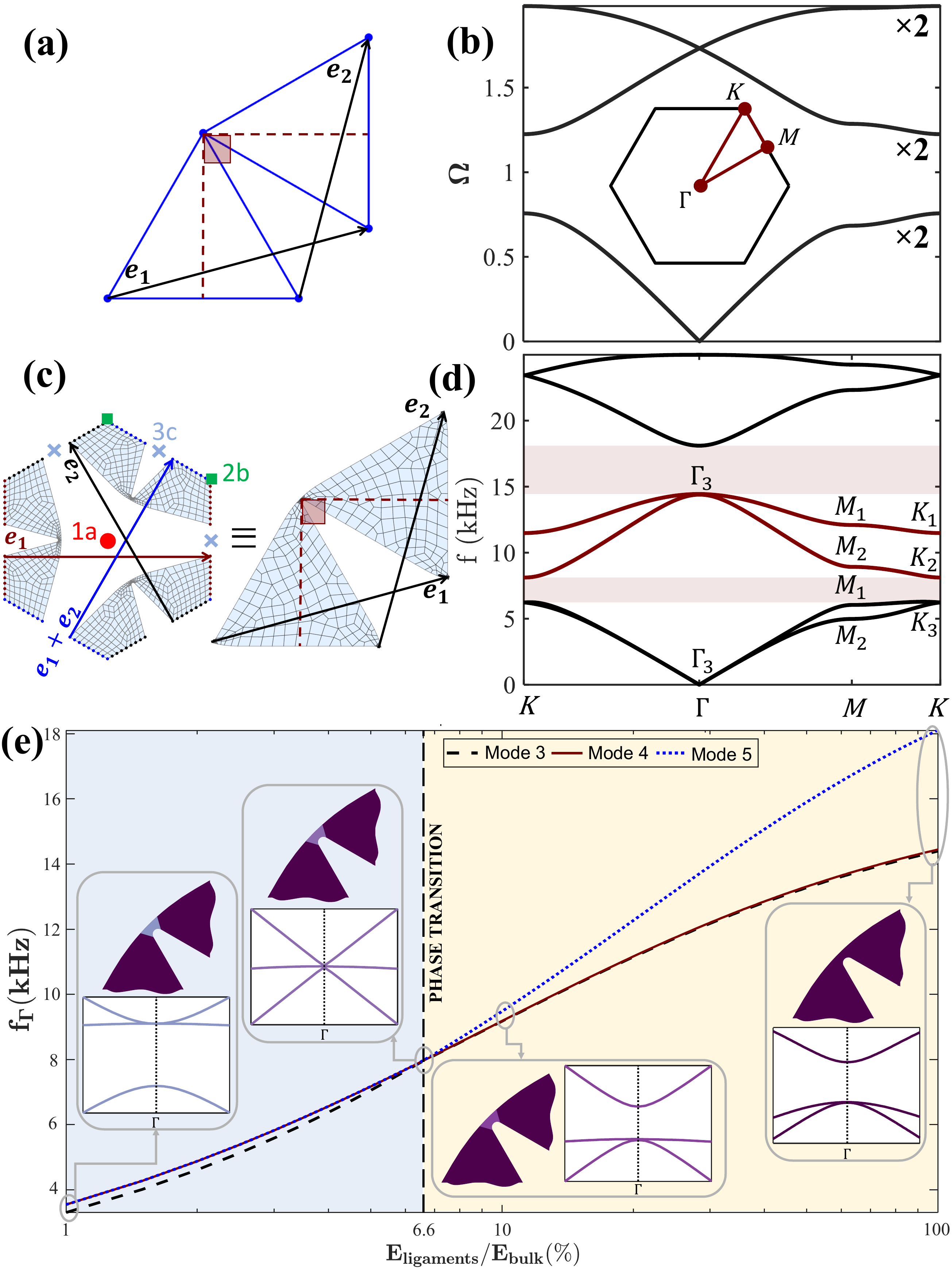}
\caption{\label{fig:model}(a) Geometry of ideal self-dual twisted kagome unit cell. (b) Corresponding band diagram with first BZ shown in inset. (c) Wigner-Seitz unit cell (left) of structural self-dual twisted kagome lattice, labeled with red dot, green square and blue cross markers representing $1a$, $2b$, and $3c$ Wyckoff positions, respectively, and its equivalent conventional unit cell (right). (d) Corresponding band diagram with emerging isolated modes 3 and 4 highlighted in maroon. The labels denote irrep at HSPs for the first four bands. The corresponding mode shapes are shown in SM.\ref{sec.1}) (e) Effects of the ligaments softening $(r=E_{ligaments}/E_{bulk})$ on the frequency f$_{\Gamma}$ of the HSP $\Gamma$ for bands 3-5, with insets showing the neighborhood of the hinge and the corresponding band diagram near $\Gamma$, color-coded proportionally to the softness ratio.}
 \end{figure}
\begin{figure*}[t]
\includegraphics[width=\textwidth]{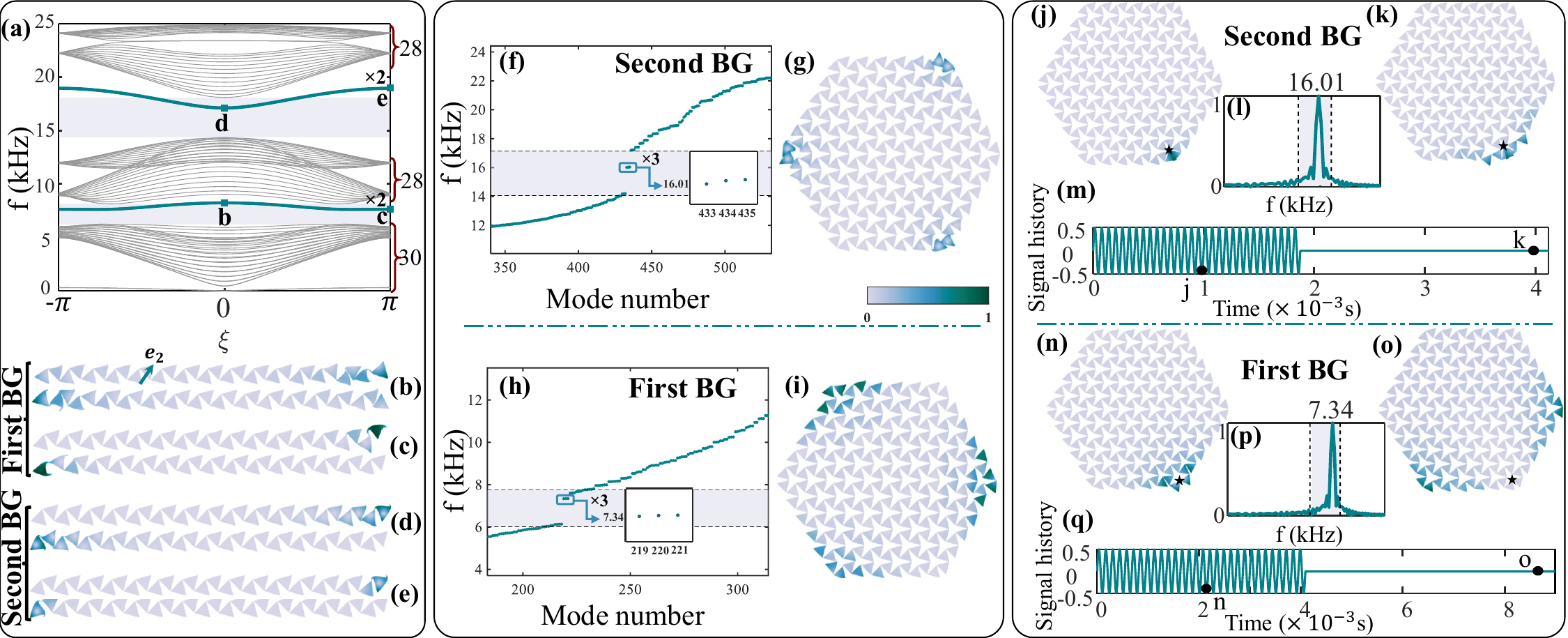}
\caption{\label{fig:corner}(a) Band diagram of a 15-cell structural self-dual twisted kagome super-cell with Bloch-periodic boundary condition along $\mathbf{e}_2$ and open boundary condition along $\mathbf{e}_1$.
(b-e) Mode shapes of degenerate edge mode pairs, sampled at $\xi=0,\pi$ and highlighted with green markers, within the first (b-c) and second (d-e) BG. (f-g) Eigenfrequencies (f) in an interval encompassing the second BG, with three degenerate modes at f=16.01 kHz, highlighted in the inset, and corresponding corner mode (from linear superposition of the degenerate modes) shown in (g). (j-m) Snapshots of wavefields for narrow-band burst excitation with carrier frequency at $\sim$16.01 kHz in the second BG (signal in (m) and corresponding spectrum in (l)), during the 30-cycle energy pumping stage (j) and relaxation time (k) of the excitation. The wavefields suggest strong localization of the second BG corner modes. (h-i),(n-q) Same quantities discussed in (f-g),(j-m) for corner modes in the first BG at 7.34 kHz. The wavefields suggest weak corner localization promoted by activation of edge modes. Colors are proportional to the in-plane displacement normalized by the maximum value.}
\end{figure*}
\begin{figure*}[t]
\includegraphics[width=\textwidth]{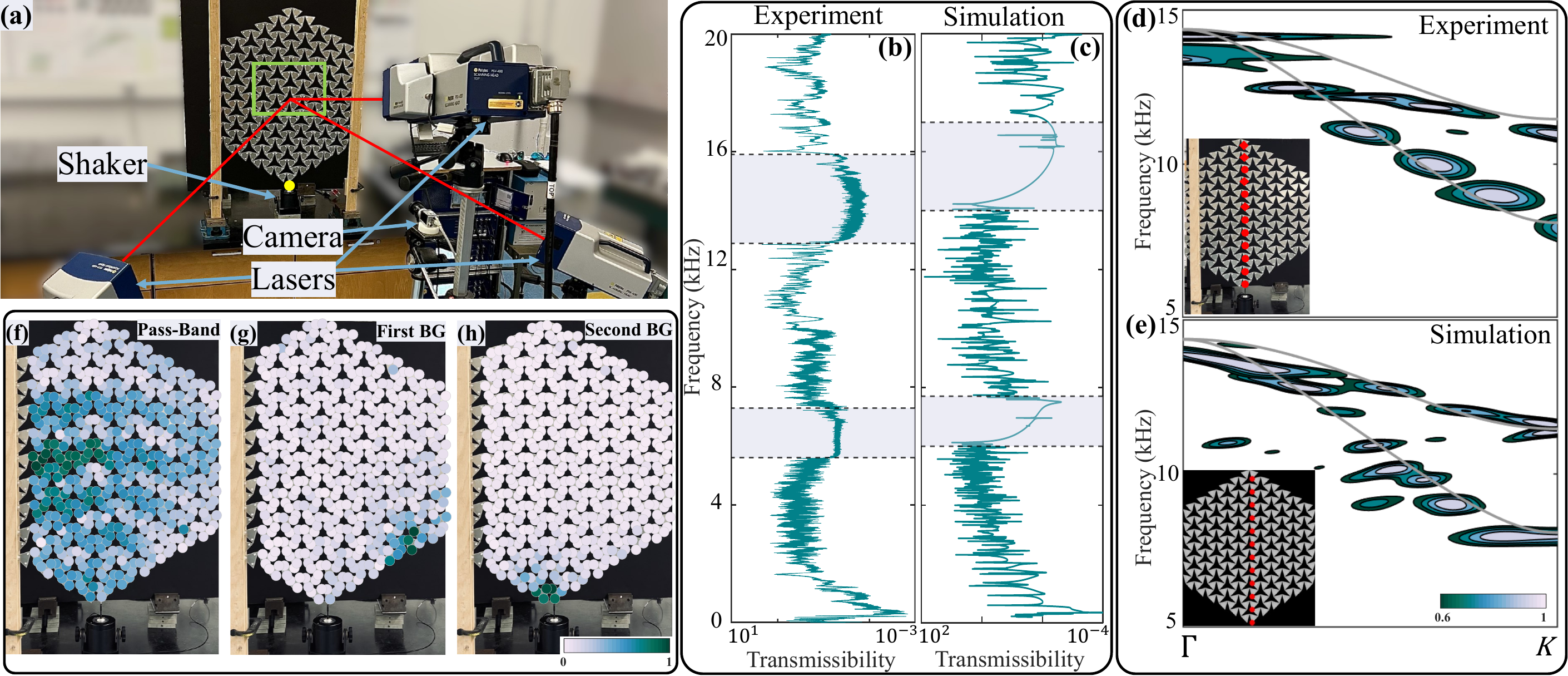}
\caption{\label{fig:exp}(a) Experimental setup for 3D SLDV testing of structural self-dual twisted kagome hexagonal prototype. (b-c) Experimental (b) and simulation (c) transmissibility curve with regions of attenuation highlighted in light purple. (d-e) Optical mode reconstruction from experimental (d) and simulation (e) data; the reconstruction involves collecting time histories at the sampling points marked by red dots in the insets and subjecting them to 2D-DFT, repeating the procedure for several carrier frequencies in the interval of interest. The 2D-DFT yields spectral amplitude contours, here superimposed to the band diagram to highlight modal activation. (f-h) Snapshots of the wavefields induced through burst excitations within the pass-band (f), and at the frequencies of the first (g) and second (h) BGs corner modes, respectively, with colorbar referring to the velocity amplitude. 2D-DFT and wavefield data are normalized by the highest value 
and proportionally color-coded.}
\end{figure*}

In this Letter, we delve deeper into the dynamics of self-dual kagome lattices in search for additional emerging behavior that can be linked to their topology. In the vein of the above-mentioned discourse on structural lattices, 
 \sg{we study the 
 case where the cell is modeled as a 2D elastodynamic continuum, which yields a dramatic reconfiguration of the BandGap (BG) landscape compared to the ideal case. Our main goal is to determine whether this transition leads to the emergence of new phenomena rooted in topology. Specifically, using Topological Quantum Chemistry (TQC) arguments, we aim at documenting the emergence of fragile topological states at finite frequencies, an extremely rare occurrence in phononic media, as detailed in \cite{https://doi.org/10.48550/arxiv.2211.11776}. As a byproduct of the band reconfiguration, we also report the availability of two spectrally distinct sets of corner modes 
 characterized by different degrees of robustness.}

Fig.~\ref{fig:model}(a) shows the unit cell of a twisted kagome lattice comprising two $90^{\circ}$-rotated equilateral triangles, a configuration known as self-dual \cite{Fruchart2020}, with $\mathbf{e}_1$ and $\mathbf{e}_2$ denoting the primitive lattice vectors. Here, the lattice consists of rods supporting only tension/compression connected with ideal hinges that allow free rotation. The corresponding band diagram, shown in Fig.~\ref{fig:model}(b), features a two-fold degenerate spectrum over the entire BZ, shown in the inset, with three pairs of overlapping bands and a double Dirac cone at the High Symmetry Point (HSP) $\Gamma$ between bands 3-6. The frequency is normalized as $\Omega = \omega/\omega_{\circ}$, where $\omega_{\circ} = \pi/L \sqrt{E/\rho}$ is the first natural frequency of a rod of length $L$, Young’s modulus $E$ and density $\rho$. Fig.~\ref{fig:model}(c) documents the transition to a structural lattice configuration, in which the triangles are elastic domains and the role of the hinges is played by finite-thickness ligaments. 
These changes induce a profound reconfiguration of the band structure (Fig.~\ref{fig:model}(d)), whereby (\romannumeral 1) the two-fold degenerate spectrum is lifted, (\romannumeral 2) a second BG is opened between bands 4 and 5, leading to the appearance of two isolated finite frequency modes, and (\romannumeral 3) a quadratic band crossing is observed at the HSP $\Gamma$, between the third and the fourth bands. We also study the evolution of the band diagram upon progressive softening of the hinges in order to reveal the existence of possible phase transitions. We introduce softening by reducing the Young's modulus of the material in the immediate neighborhood of the hinges, as captured by the ratio $r=E_{ligaments}/E_{bulk}$ where, for instance, $r=1\%$ yields very soft hinges while $r=100\%$ returns the mono-material cell. Specifically, here we monitor the frequencies f$_{\Gamma}$ of bands 3-5 at the HSP $\Gamma$, whose evolution with $r$ is illustrated in Fig.~\ref{fig:model}(e). We find that, as the hinge properties evolve, the gap closes and reopens at a certain value of the control parameter (here $\backsim 6.6\%$), with the quadratic crossing migrating from below to above the gap, which unequivocally denotes a phase transition. The insets show, for a few selected configurations, the immediate neighborhood of the hinge along with the corresponding band diagram (zoom on the neighborhood of the HSP $\Gamma$) with hinge and bands color-coded proportionally to $r$. 
Interestingly, at the critical point, the band crossing becomes linear and exhibits a three-fold degeneracy with an almost flat band (Dirac-like cone). Recently, an analogous linear regime in the neighborhood of Dirac points has been demonstrated for periodic origami using a dynamic homogenization framework \cite{OUDGHIRIIDRISSI2022115386}.  

We now claim that the nature of the above-mentioned phase transition is topological. Specifically, reopening the BG induces fragile topological states within bands 3 and 4 protected by three-fold rotation symmetry $C_3$. This can be shown using the recently developed method of symmetry indicators~\cite{po2017symmetry,khalaf2018symmetry}, band structure combinatorics~\cite{kruthoff2017topological} and TQC~\cite{bradlyn2017topological,cano2018building}. This method provides a full classification of topological states protected by spatial symmetries, and can detect any such topological states by investigating how the eigenfunctions of the bands transform under symmetries, i.e., the representation labels of the eigenfunctions, at the HSPs of the BZ. Conveniently, all possible representation labels that a set of bands generated by SLWF can feature are tabulated in the Bilbao Crystallography Server (BCS)~\cite{aroyo2011crystallography,aroyo2006bilbaoI,aroyo2006bilbaoII,vergniory2017graph,elcoro2017double}. We note that our system falls under $G = p31m$ wallpaper group. The HSPs $\Gamma$, $K$ and $M$ (inset of Fig.~\ref{fig:model}(b)) have little co-groups $C_{3v}$, $C_{3v}$ and $C_{s}$, respectively. The eigenfunctions at these HSPs transform under the representations of the corresponding little co-group. From FEM, we find that degenerate eigenfunctions at $\Gamma$ point of bands 3-4 transform under 2d irreducible representation (irrep) $\Gamma_3$, the eigenfunctions at $M$ transform under 1d irreps $M_2$ and $M_1$, and the eigenfunctions at $K$ transform under 1d irreps $K_2$ and $K_1$, as denoted in Fig.~\ref{fig:model}(d) (see also SM.\ref{sec.1} for corresponding eigenfunctions). BCS tables reveal that these irreps cannot be represented by any SLWF, which implies that these two bands are topological~\cite{bradlyn2017topological}\sg{; as} a result, the change in behavior observed at $r=6.6\%$ in Fig.~\ref{fig:model}(e) can be qualified as a topological phase transition. \sg{Specifically,} the irreps of bands 3-4 are consistent with the formal \textit{difference}:
\begin{equation}
   (^1E^2E\uparrow G)_{2b} \ominus (E\uparrow G)_{1a},
\end{equation}
where $^1E^2E$ (and $E$) represents two SLWFs with angular momentum 1 ($p_x$ and $p_y$ type orbitals), where $1a$ and $2b$ are Wyckoff positions shown in Fig.~\ref{fig:model}(c). In words, this means that irrep labels for bands 3 and 4 are consistent with the \textit{difference} between $p_x-p_y$ type orbitals at position $2b$ (see Fig.~\ref{fig:model}(c); position $2b$ has multiplicity 2 in the unit cell, so in total 4 orbitals in the unit cell) and $p_x-p_y$ type orbitals at position $1a$ (position $1a$ has multiplicity 1 in the unit cell, in total 2 orbitals in the unit cell). 
Furthermore, since the irrep labels of the bands 1-2  are $\Gamma_3 - M_1\oplus M_2 - K_3$ and are consistent with the SLWFs $(E\uparrow G)_{1a}$ (which renders bands 1-2 topologically trivial), together bands 1-4 can be represented by SLWFs $(^1E^2E\uparrow G)_{2b}$, meaning that bands 1-4 together are topologically trivial. This is the distinguishing characteristic of fragile topological bands - addition of trivial bands to them render the topology trivial~\cite{song2020twisted}. We also verify the topological non-triviality of bands 3-4 by calculating the winding of hexagonal Wilson Loop (WL)~\cite{bradlyn2019disconnected} (see SM.\ref{sec.2} for details of WL calculation). Importantly, here the fragile topology of bands 3-4 is protected by $C_3$ (that this fragile topology only depends on $C_3$, not the mirrors can be understood from the fact that, if we break the mirror symmetry perturbatively such that bands 3-4 are still isolated from other bands, the irrep labels at the HSPs would be $\Gamma_2\oplus\Gamma_3-2M_1-2K_1$, which also correspond to fragile topological bands~\cite{song2020twisted}). 
\sg{It is worth noting that the fragile topology of bands 3-4 can be detected just from the band connectivity, namely from the fact that there is a degeneracy at $\Gamma$ and no degeneracy at $K$ -something that trivial bands (that can be represented by SLWF) cannot display, as detailed in SM.\ref{sec.3}. 
This makes the detection of $C_3$ symmetry protected fragile bands in wallpaper group $p31m$ considerably easier than any other type of fragile bands where further explicit calculations are required to confirm nontrivial topology.}

We now shift our attention to the two BGs bounding \sg{bands 3-4}, 
looking for any edge and corner modes comprised therein.  
\sg{We} first perform a super-cell analysis on a 15-cell super-cell modeled with the same finite element discretization used for the unit cell analysis. The resulting band diagram is plotted in Fig.~\ref{fig:corner}(a). 
We observe a pair of degenerate bands in each BG (shaded regions) indicated by green lines. The super-cell mode shapes at $\xi=0$ and $\xi=\pi$ within the first and the second BGs (marked by the green markers) are displayed in Figs.~\ref{fig:corner}(b-c) and (d-e), respectively. The high decay rate and the \sg{occurrence} of localization at two opposite edges qualify these branches as non-polarized edge modes, as expected for a twisted kagome lattice (see SM.\ref{sec.4} for details). 
We then calculate eigenfrequencies and mode shapes for a finite hexagon-shaped domain as shown in Figs.~\ref{fig:corner}(g) and (i). The eigenfrequencies in intervals encompassing the two BGs are shown in Figs.~\ref{fig:corner}(f) and \ref{fig:corner}(h). The insets zoom in on two sets of three degenerate modes inside the second (f =16.01 kHz) and first (f =7.34 kHz) BG; the corresponding mode shapes depicted in Figs.~\ref{fig:corner}(g) and \ref{fig:corner}(i), obtained by superposition of the degenerate modes, reveal localization of deformation at the corners of the hexagon, which qualifies these as corner modes and the lattice as 2D second-order TI. 
Interestingly, the two sets of mode shapes feature localization at alternating corners and have distinct morphological characteristics. Specifically, corner modes appearing in the first BG have lower decay rate, likely due to their higher proximity to (and contamination from) the bulk band and the edge modes observable in the top region of the gap. Conversely, the second BG corner modes show stronger localization. Additional evidence of these differences is found via full-scale transient simulations, in which we excite the bottom-right corner (marked by a black star) 
with 30-cycle narrow-band tone burst force excitations with carrier frequencies at 16.01 and 7.34 kHz, corresponding to the second BG and first BG corner modes, respectively (time histories and spectra are depicted in Figs.~\ref{fig:corner}(m),(q) and Figs.~\ref{fig:corner}(l),(p)). For each carrier, two snapshots of the resulting wavefields are shown in Figs.~\ref{fig:corner}(j),(k) for the second BG and Figs.~\ref{fig:corner}(n),(o) for the first BG. For the second BG, we observe localization in the neighborhood of the excitation point, matching the mode shape pattern in Fig.~\ref{fig:corner}(g) and confirming the notion that the second BG corner modes minimally leak into edge and bulk modes. In contrast, the wavefield for the first BG corner modes shows opposite dynamics: while in the early time snapshot of Fig.~\ref{fig:corner}(n) some degree of localization is established around the excitation point, the signal travels along the edge and eventually migrates to the adjacent corner, as seen in Fig.~\ref{fig:corner}(o). This transfer can be explained by the combination of three factors: the lack of a corner mode for the excited corner in the first BG, the availability of a corner mode, at that same frequency, in the adjacent corner, and the spectral proximity of the signal carrier to prominent edge modes. 
\sg{We note that, while no formal connection can be claimed between fragile topological states and edge/corner modes in the absence of a proper bulk-edge correspondence for fragile topology, the availability of localized modes is in itself a byproduct (of this band structure) of practical relevance, e.g., for vibration isolation and harvesting applications.} 

We substantiate our theoretical findings - emergence of fragile topological states and establishment of corner modes - through laser vibrometry 
experiments on a physical prototype, see setup in Fig.~\ref{fig:exp}(a). All details on fabrication, vibrometer specifications and setup are provided in SM.\ref{sec.5}-SM.\ref{sec.7}.  First, we want to demonstrate experimentally the opening of the second BG, which is absent in the ideal self-dual kagome lattice, as a result of the finite-thickness hinges of the structural metamaterial. To this end, we prescribe a broadband pseudo-random excitation at the point marked by the yellow dot (Fig.~\ref{fig:exp}(a)), measure the in-plane velocity at the designated sampling points inside the green box (Fig.~\ref{fig:exp}(a)), and normalize the average value by that of the excitation point to construct a curve of transmissibility versus frequency, plotted in Fig.~\ref{fig:exp}(b). An analogous curve, plotted in Fig.~\ref{fig:exp}(c), is obtained via full-scale steady-state simulation 
under sustained harmonic excitation. 
Both curves, in reasonable agreement, feature two distinct regions of attenuation, highlighted in light purple, supporting the existence of finite-frequency isolated band(s) in the mid-frequency spectrum. The attenuation intervals are in satisfactory agreement with the BGs predicted via Bloch analysis (Fig.~\ref{fig:model}(d)). See also SM.\ref{sec.8} for details on the discrepancies between experiments and theory. 

We now seek evidence of the fragile topological nature of the intermediate bands by experimentally reconstructing the morphological characteristics of the bands. To this end, we collect the time histories of the lateral in-plane velocity at evenly distanced points along the $\Gamma$-$K$ direction (red dots in the insets of the Figs.~\ref{fig:exp}(d),(e), for experiment and simulation, respectively) for several tone burst excitation signals with different carrier frequencies within the pass-band regions. Subsequently, we perform 2D Discrete Fourier Transform (2D-DFT) on each spatio-temporal data set and we aggregate all the resulting spectral amplitude contours and superimpose them on the band diagram  (see SM.\ref{sec.9} for full depiction). Overall the contours of the bursts populate the spectral plane in a way that follows the morphological attributes of the band diagram. Focusing on the mid-frequency range, the peculiar shape of the third and forth modes associated with fragile topology, featuring the bands crossing at $\Gamma$ without retouching elsewhere at the boundary of the BZ [or at the other HSPs] is captured perfectly. Finally, we conduct three experiments with three distinct tone bursts (applied at the bottom corner) at three carrier frequencies falling in the acoustic pass-band, first BG and second BG. The objective here is to reconstruct experimentally the wave propagation patterns established in different frequency regimes and document the establishment of corner modes. In Fig.~\ref{fig:exp}(f) a carrier frequency of $\sim$3 kHz generates a wavefield that propagates through the bulk, which is the signature of pass-band behavior. For an excitation in the first BG (Fig.~\ref{fig:exp}(g)), we observe wave propagation along the edge and accumulation at the next available corner. The behavior is consistent with the localized modal landscape in the first BG, in which the corner modes do not appear at the set of corners that include the excitation point and, moreover, well-defined edge modes exist in the gap. In contrast, for excitation in the second BG, we observe a highly localized and persistent corner mode (Fig.~\ref{fig:exp}(h)), even after long relaxation times for the bursts.


P.A. acknowledges the support of the UMN CSE Graduate Fellowship. S.G. acknowledges support from the National Science Foundation (grant CMMI-2027000). S.S. and K.S. acknowledge the support from the Office of Naval Research (grant MURI N00014-20-1-2479).


\bibliographystyle{apsrev4-1}
\bibliography{ref}

\onecolumngrid

\makeatletter
\renewcommand \thesection{S-\@arabic\c@section}
\renewcommand\thetable{S\@arabic\c@table}
\renewcommand \thefigure{S\@arabic\c@figure}
\renewcommand \theequation{S\@arabic\c@equation}
\makeatother
\setcounter{equation}{0}  
\setcounter{figure}{0}  
\setcounter{section}{0}  

\section*{\Large\bf Supplemental Material}
\maketitle
\section{Mode shapes of the lowest four bands at the High Symmetry Points (HSP) of the Brillouin Zone (BZ)}\label{sec.1}
\begin{figure}[h!]
    \centering
    \includegraphics[width = 0.7\textwidth]{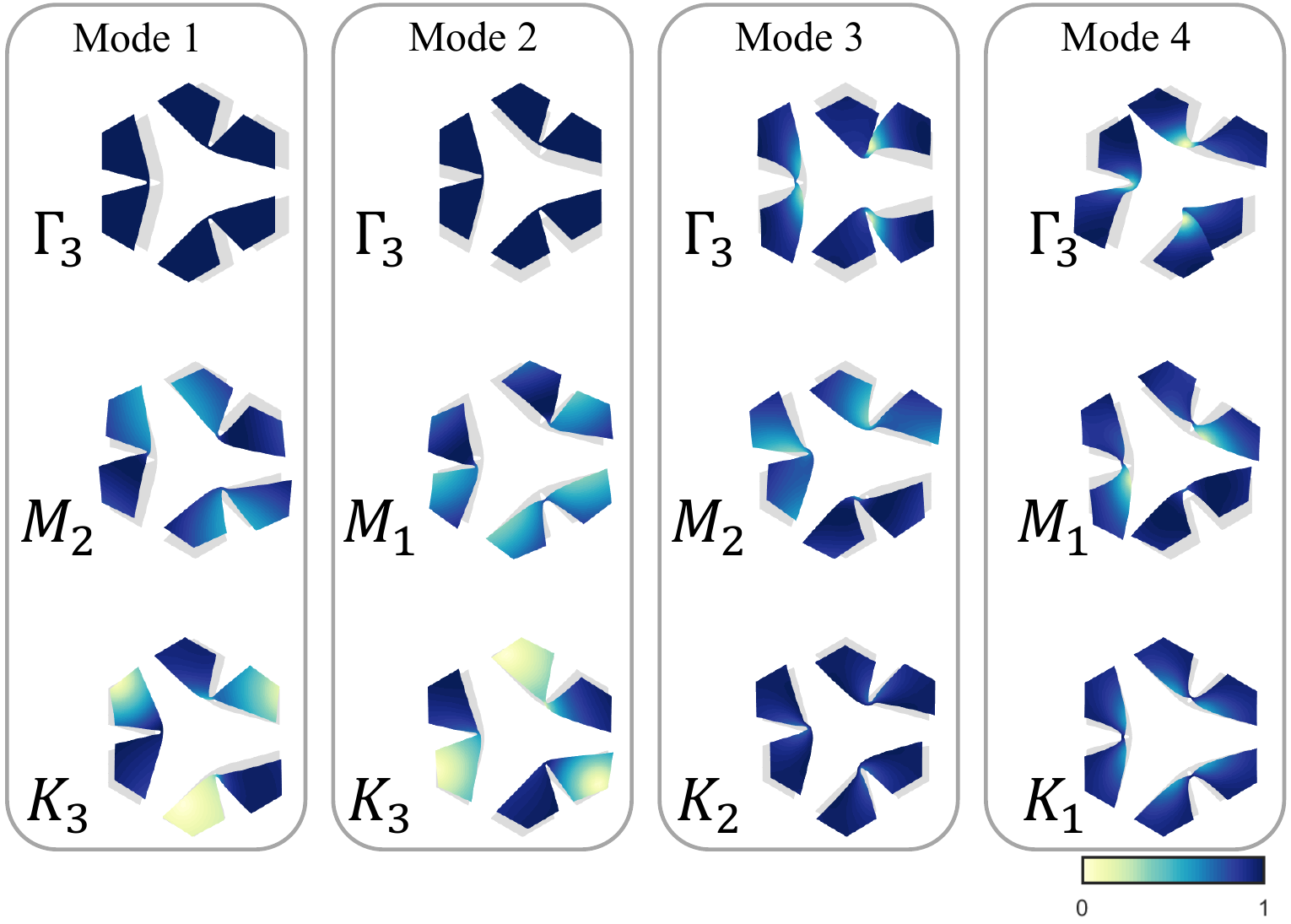}
    \caption{Deformation of the unit cell corresponding to the eigenfunctions at HSPs $\Gamma$ (first row), $M$ (second row) and $K$ (third row). Mode numbers are counted from the lowest band upward. The irreducible representation (irrep) labels are marked along with the mode shapes.}
    \label{fig:ModesWS}
\end{figure}

In Fig.~\ref{fig:ModesWS}, we show the mode shapes of the the lowest four bands at the HSPs. The lowest two modes at $\Gamma$ are just the two uniform translations. Since translation transforms as a vector, they form $\Gamma_3$ representation (see the Bilbao Crystallography Server (BCS) for the notation~\cite{aroyo2011crystallography1,aroyo2006bilbaoI1,aroyo2006bilbaoII1}). Modes 3 and 4 at $\Gamma$ are even and odd under horizontal mirror (passing through the center of the unit cell) reflection, but transform into each other under $C_3$, hence they also form $\Gamma_3$ representation. The modes at $M$ each form the basis of $M_1$ or $M_2$ representation depending on if they are even or odd under the mirror parallel to $(-1/2,\sqrt{3}/2)$ direction and passing through the center of the unit cell. The lowest two modes at $K$ transform into each other under mirror reflection and $C_3$, hence they form $K_3$ representation. Mode 3 (4) at $K$ transforms into itself under $C_3$ but is odd (even) under the horizontal mirror; hence it forms $K_2$ ($K_1$) representation.

\section{Wilson Loop (WL) Calculation}\label{sec.2}
\begin{figure}[b]
    \centering
    \includegraphics[width = 0.7\textwidth]{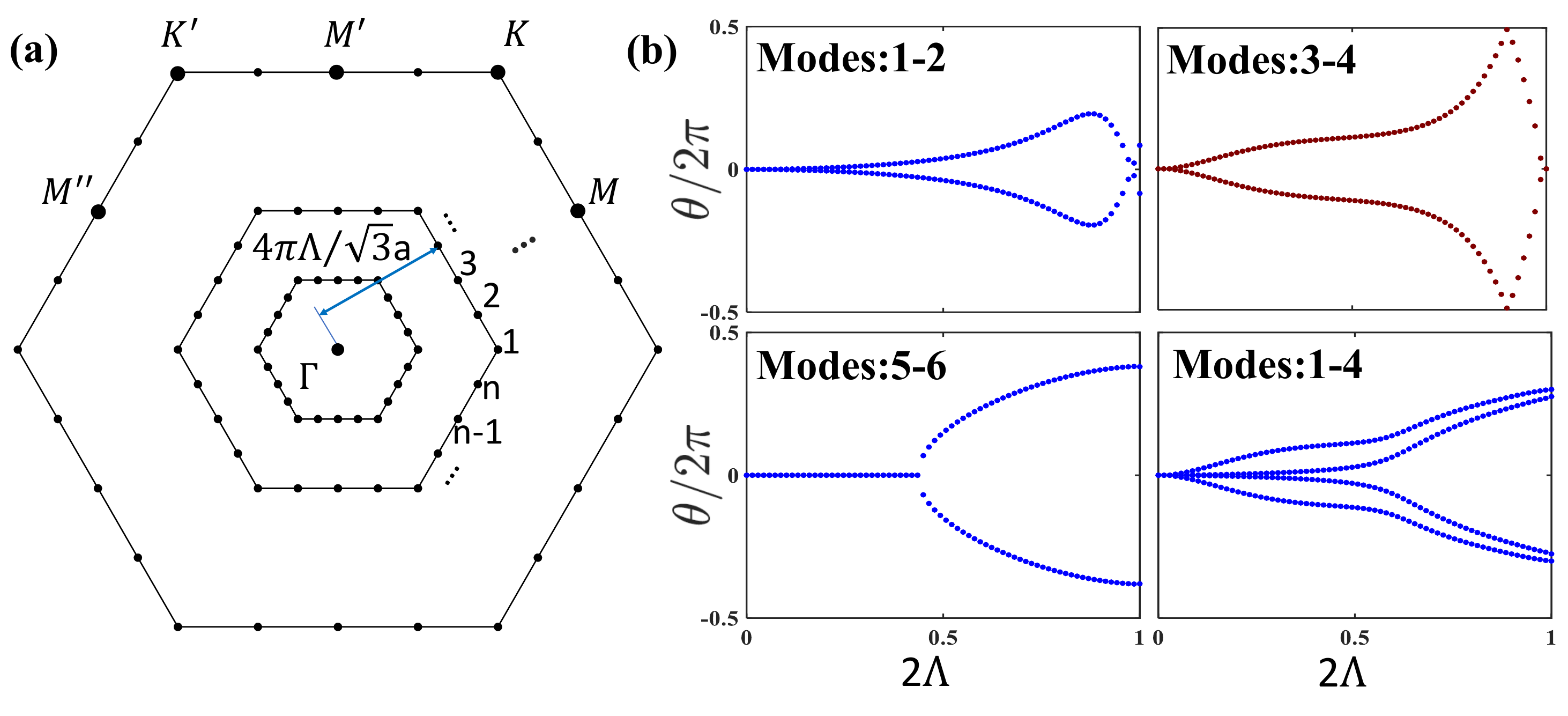}
    \caption{(a) Descritized hexagonal loops using for WL calculation. The outer loop represents BZ and the distance of each loop from the HSP $\Gamma$ is $4\pi\Lambda/\sqrt{3}a$. (b) WL of the first-second bands (top left), third-fourth bands (top right), fifth-sixth bands (bottom left), first-fourth bands (bottom right).}
    \label{fig:WL}
\end{figure}

In the main text we established the fact that the fragile topology of bulk bands 3-4 is protected by three-fold rotation symmetry $C_3$. Here, we will show that the nontrivial topology of the fragile bands can be captured by the winding of a hexagonal WL (see Fig.~\ref{fig:WL}(a)). Hexagonal WL was introduced in~\cite{bradlyn2019disconnected1} (for a good review on WLs see~\cite{alexandradinata2014wilson1}) and was used on a tight binding model with $p3m1$ symmetry. Here, we have $p31m$ space group, but we show analytically (slightly modifying the expression in~\cite{bradlyn2019disconnected1} to adapt it to $p31m$ space group) that the eigenvalues of the hexagonal WL for bands 3-4 still winds. To do so, we start by setting the lattice vectors and reciprocal lattice vectors to be
\begin{equation}
\begin{split}
\mathbf{e}_1 &= a\mathbf{\hat{x}}\\
\mathbf{e}_2 &=a\left(- \frac{1}{2}\mathbf{\hat{x}}+\frac{\sqrt{3}}{2}\mathbf{\hat{y}}\right)\\
\mathbf{g}_1 &= \frac{4\pi}{\sqrt{3}a}\left(\frac{\sqrt{3}}{2}\mathbf{\hat{x}} + \frac{1}{2}\mathbf{\hat{y}}\right)\\
\mathbf{g}_2 &= \frac{4\pi}{\sqrt{3}a}\mathbf{\hat{y}}\\
\end{split}
\end{equation}
such that $\mathbf{e}_i \cdot \mathbf{g}_j = 2\pi \delta_{ij}$. Written in the basis of $\mathbf{g}_1$ and $\mathbf{g}_2$, the HSPs are given by $\Gamma = (0,0)$, $M = (1/2,0)$, $M' = (0,1/2)$, $M'' = (-1/2,1/2)$, $K = (1/3,1/3)$, $K' =(-1/3,2/3)$. As shown in Fig.~\ref{fig:WL}(a), the hexagonal WL at a distance $\Lambda$ from the $\Gamma$ point is defined as:
\begin{equation}
\begin{split}
W_h(\Lambda) = &W_{(4 \Lambda/3,-2 \Lambda/3)\leftarrow(2 \Lambda/3,-4 \Lambda/3)}W_{(2 \Lambda/3,-4 \Lambda/3)\leftarrow(-2 \Lambda/3,-2 \Lambda/3)}W_{(-2 \Lambda/3,-2 \Lambda/3)\leftarrow(-4 \Lambda/3,2 \Lambda/3)} \times\\
&\;\;W_{(-4 \Lambda/3,2 \Lambda/3)\leftarrow(-2 \Lambda/3,4 \Lambda/3)}W_{(-2 \Lambda/3,4 \Lambda/3)\leftarrow(2 \Lambda/3,2 \Lambda/3)}W_{(2 \Lambda/3,2 \Lambda/3)\leftarrow (4 \Lambda/3,-2 \Lambda/3)},
\end{split}
\end{equation}
where $W_{\mathbf{k}_2\leftarrow\mathbf{k}_1}$ is defined on a \textit{straight line} joining $\mathbf{k}_1$ and $\mathbf{k}_2$ as
\begin{equation}
\begin{split}
W_{\mathbf{k}_2\leftarrow\mathbf{k}_1}^{mn} &= \langle u_{\mathbf{k}_2}^m|  \prod_\mathbf{k}^{\mathbf{k}_2\leftarrow\mathbf{k}_1} P(\mathbf{k}) |u_{\mathbf{k}_1}^n\rangle,\\
P^{ab}(\mathbf{k}) &=\sum_{i \in \text{bands being considered}} | u_{\mathbf{k}}^a\rangle  \langle u_{\mathbf{k}}^b |,
\end{split}
\end{equation}
where $| u_{\mathbf{k}}^a\rangle $ is the $a^{\tiny\text{th}}$ eigenvector (of the modified dynamical matrix $\tilde{D}(\mathbf{k}) = M^{-1/2}D(\mathbf{k})M^{-1/2}$, where $M$ is mass matrix) at momentum $\mathbf{k}$, the ordered product is being taken over $N$ points on the straight line joining $\mathbf{k}_1$ and $\mathbf{k}_2$ with some discretization of the line into $N$ points (see Fig.~\ref{fig:WL}(a)), and $P(\mathbf{k})$ is the projection operator onto the bands under consideration at point $\mathbf{k}$. Note that in the limit of $N\rightarrow \infty$, $W_{\mathbf{k}_2\leftarrow\mathbf{k}_1}$ is a unitary matrix satisfying~\cite{alexandradinata2014wilson1}:
\begin{equation}
\label{eq:WLProperties}
W_{\mathbf{k}_2\leftarrow\mathbf{k}_1}^{-1} = W_{\mathbf{k}_2\leftarrow\mathbf{k}_1}^\dagger =W_{\mathbf{k}_1\leftarrow\mathbf{k}_2}.
\end{equation}
Therefore, $W_h(\Lambda)$ is also unitary and its eigenvalues are of the form $e^{i \theta(\Lambda)}$. If we only consider bands  3 and 4, $W_{h,3-4}(\Lambda)$ is a $2\times 2$ matrix. In the following, using the irrep labels of band 3-4 at HSPs, we prove that $\theta_{3-4}^{(1)}(\Lambda)$ and $\theta_{3-4}^{(2)}(\Lambda)$ ($e^{i \theta_{3-4}^{(1)}(\Lambda)}$ and $e^{i \theta_{3-4}^{(2)}(\Lambda)}$ being the eigenvalues of $W_{h,3-4}(\Lambda)$) wind in the opposite direction as shown in Fig.~\ref{fig:WL}(b). 
To this end, we define a Wilson line over the path $\mathcal{P} = (-2 \Lambda/3,4 \Lambda/3) \leftarrow (2 \Lambda/3,2 \Lambda/3) \leftarrow(4 \Lambda/3,-2 \Lambda/3)$:
\begin{equation}
W_\mathcal{P}(\Lambda) = W_{(-2 \Lambda/3,4 \Lambda/3)\leftarrow(2 \Lambda/3,2 \Lambda/3)}W_{(2 \Lambda/3,2 \Lambda/3)\leftarrow (4 \Lambda/3,-2 \Lambda/3)}.
\end{equation}
Then, the following is true:
\begin{equation}
\begin{split}
W_h(\Lambda) &= W_{C_3^2\mathcal{P}}(\Lambda)W_{C_3\mathcal{P}}(\Lambda) W_\mathcal{P}(\Lambda)\\
 &= C_3^2W_{\mathcal{P}}(\Lambda)C_3^{-2}C_3W_{\mathcal{P}}(\Lambda) C_3^{-1}W_\mathcal{P}(\Lambda)\\
 &= C_3^{-1}W_{\mathcal{P}}(\Lambda)C_3^{-1}W_{\mathcal{P}}(\Lambda) C_3^{-1}W_\mathcal{P}(\Lambda)\\
 &= (C_3^{-1}W_{\mathcal{P}}(\Lambda))^3,
\end{split}
\end{equation}
where $C_3\mathcal{P}$ is the path obtained by rotating $\mathcal{P}$ counterclockwise by an angle $2\pi/3$ around the $\Gamma$ point, and we used $C_3^2 = C_3^{-1}$. Defining $W_3(\Lambda) \equiv C_3^{-1}W_{\mathcal{P}}(\Lambda)$, we have $W_{h}(\Lambda) = (W_3(\Lambda))^3$. Since $W_{\mathcal{P}}(\Lambda)$ has matrix elements between $|u_{(4\Lambda/3,-2\Lambda/3)}\rangle$ and $|u_{(-2\Lambda/3,4\Lambda/3)}\rangle$, and $C_3^{-1}$ has matrix elements between $|u_\mathbf{k}\rangle$ and $|u_{C_3^{-1}\mathbf{k}}\rangle$, the matrix elements of $W_3(\Lambda)$ are
\begin{equation}
W_3^{mn}(\Lambda) = \langle u_{(4\Lambda/3,-2\Lambda/3)}^m| C_3^{-1}\left(\sum_{l \in \text{bands being considered}} |u_{(-2\Lambda/3,4\Lambda/3)}^l\rangle \langle u_{(-2\Lambda/3,4\Lambda/3)}^l| \right)W_\mathcal{P} |u_{(4\Lambda/3,-2\Lambda/3)}^n\rangle,
\end{equation}
which implies that under some unitary gauge transformation of the eigenvectors $ |u_{(4\Lambda/3,-2\Lambda/3)}^m\rangle \rightarrow G^{mn} |u_{(4\Lambda/3,-2\Lambda/3)}^n\rangle $, $W_3$ transforms covariantly (i.e., $W_3 \rightarrow G W_3 G^\dagger$); as a result the eigenvalues of $W_3(\Lambda)$ are gauge invariant.

Next we show that due to mirror symmetry $m_y$ ($m_y(x,y) = (x,-y)$), the eigenvalues of $W_3(\Lambda)$ come in pairs $e^{\pm i \theta_3 (\Lambda)}$ and consequently, the eigenvalues of $W_h(\Lambda)$ come in pairs $e^{\pm i 3\theta_3 (\Lambda)}$. To show this, we note that
\begin{equation}
\begin{split}
m_y W_3(\Lambda)m_y^{-1} &= m_y C_3^{-1}W_{\mathcal{P}}(\Lambda)m_y^{-1}\\
&= C_3 m_y W_{\mathcal{P}}(\Lambda)m_y^{-1}\\
&= C_3 m_y W_{(-2 \Lambda/3,4 \Lambda/3)\leftarrow(2 \Lambda/3,2 \Lambda/3)}W_{(2 \Lambda/3,2 \Lambda/3)\leftarrow (4 \Lambda/3,-2 \Lambda/3)}m_y^{-1}\\
&= C_3 m_y W_{(-2 \Lambda/3,4 \Lambda/3)\leftarrow(2 \Lambda/3,2 \Lambda/3)}m_y^{-1}m_yW_{(2 \Lambda/3,2 \Lambda/3)\leftarrow (4 \Lambda/3,-2 \Lambda/3)}m_y^{-1}\\
&=C_3 W_{(-2 \Lambda/3,-2 \Lambda/3)\leftarrow(2 \Lambda/3,-4 \Lambda/3)} W_{(2 \Lambda/3,-4 \Lambda/3)\leftarrow (4 \Lambda/3,-2 \Lambda/3)}\\
&= (C_3 W_{(-2 \Lambda/3,-2 \Lambda/3)\leftarrow(2 \Lambda/3,-4 \Lambda/3)} C_3^{-1}) (C_3 W_{(2 \Lambda/3,-4 \Lambda/3)\leftarrow (4 \Lambda/3,-2 \Lambda/3)}C_3^{-1})C_3\\
&= W_{ (4 \Lambda/3,-2 \Lambda/3)\leftarrow(2 \Lambda/3,2 \Lambda/3)} W_{(2 \Lambda/3,2 \Lambda/3)\leftarrow(-2 \Lambda/3,4 \Lambda/3)} C_3\\
&= (C_3^{-1} W_{(-2 \Lambda/3,4 \Lambda/3)\leftarrow(2 \Lambda/3,2 \Lambda/3)}W_{(2 \Lambda/3,2 \Lambda/3)\leftarrow (4 \Lambda/3,-2 \Lambda/3)})^{-1}\\
&= W_3^{-1}(\Lambda),
\end{split}
\end{equation}
where going from first to second equality, we used $m_y C_3^{-1} m_y^{-1} =C_3$, and in various places we used the fact the properties of $W$ in Eq.~\eqref{eq:WLProperties}. This implies that the set of eigenvalues of $W_3(\Lambda)$ and $W_3^{-1}(\Lambda)$ are same, i.e., the eigenvalues of $W_3(\Lambda)$ come in pairs $e^{\pm i \theta_3 (\Lambda)}$. 

Now, we specialize to bands 3-4 to check the eigenvalues of $W_{3,3-4}(\Lambda)$ and $W_{h,3-4}(\Lambda)$ for different values of $\Lambda$. At $\Lambda = 0$, we have
\begin{equation}
W_3^{mn}(0) = \langle u_{(0,0)}^m| C_3^{-1}  |u_{(0,0)}^n\rangle,
\end{equation}
which is the representation matrix of $C_3^{-1}$ in the basis of eigenvectors of bands 3-4 at $\Gamma$ point. Since the irrep label of bands 3-4 is $\Gamma_3$, the eigenvalues of $W_3(0)$ are $\omega$ and $\omega^2$ where $\omega = (-1+\sqrt{3}i)/2$. This implies $\{e^{i\theta_{3,3-4}(0)}\} = \{e^{i 2\pi/3},e^{-i 2\pi/3}\}$. Consequently, we have
\begin{equation}
W_h^{mn}(0) = \delta^{mn},
\end{equation}
and $\{e^{i\theta_{3-4}(0)}\}= \{1,1\}$. 

Things are more complicated at $\Lambda = 1/2$:
\begin{equation}
\begin{split}
W_3(1/2) &= C_3^{-1} W_{K'\leftarrow K}W_{K \leftarrow (K'+\mathbf{g}_1-\mathbf{g}_2)}\\
&= (C_3^{-1} W_{K'\leftarrow K}C_3) C_3^{-1}W_{K \leftarrow K'+\mathbf{g}_1-\mathbf{g}_2}\\
&= W_{(K'+\mathbf{g}_1-\mathbf{g}_2)\leftarrow (K-\mathbf{g}_2)} C_3^{-1}W_{K \leftarrow K'+\mathbf{g}_1-\mathbf{g}_2}\\
&= V(\mathbf{g}_1-\mathbf{g}_2) W_{K'\leftarrow (K-\mathbf{g}_1)} V^\dagger(\mathbf{g}_1-\mathbf{g}_2)  C_3^{-1}W_{K \leftarrow K'+\mathbf{g}_1-\mathbf{g}_2}\\
&= V(\mathbf{g}_1-\mathbf{g}_2) W_{K'\leftarrow (K-\mathbf{g}_1)} V^\dagger(\mathbf{g}_1-\mathbf{g}_2) C_3^{-2}  (C_3 W_{K \leftarrow K'+\mathbf{g}_1-\mathbf{g}_2}C_3^{-1}) C_3\\
&=V(\mathbf{g}_1-\mathbf{g}_2) W_{K'\leftarrow (K-\mathbf{g}_1)} V^\dagger(\mathbf{g}_1-\mathbf{g}_2) C_3^{-2} W_{K -\mathbf{g}_1\leftarrow K'}C_3\\
&=V(\mathbf{g}_1-\mathbf{g}_2) W_{K'\leftarrow (K-\mathbf{g}_1)} V^\dagger(\mathbf{g}_1-\mathbf{g}_2) C_3^{-2} W_{K -\mathbf{g}_1\leftarrow K'}C_3\\
C_3 W_3(1/2)C_3^{-1} &=  C_3V(\mathbf{g}_1-\mathbf{g}_2) W_{K'\leftarrow (K-\mathbf{g}_1)} V^\dagger(\mathbf{g}_1-\mathbf{g}_2) C_3 W_{K -\mathbf{g}_1\leftarrow K'}\\
&= B_{C_3}(K'+\mathbf{g}_1-\mathbf{g}_2) W_{K'\leftarrow (K-\mathbf{g}_1)} B_{C_3}(K -\mathbf{g}_1) W_{K -\mathbf{g}_1\leftarrow K'},
\end{split}
\end{equation}
where the matrix $V(\mathbf{g})$ at reciprocal lattice vector $\mathbf{g}$ is defined as $|u_{\mathbf{k}+\mathbf{g}}^m\rangle = V^{mn}(\mathbf{g}) |u_{\mathbf{k}}^n\rangle$ (i.e., for singly degenerate bands $V(\mathbf{g}) = e^{-i \mathbf{g}\cdot \mathbf{x}}\mathbbm{1}$), and the sewing matrix $B_{C_3}(\mathbf{k})$ is defined as $B_{C_3}(\mathbf{k}) =\langle u_{C_3\mathbf{k}}|C_3|u_{\mathbf{k}}\rangle$. Now, since the irreps at the $K$ point are $K_1$ and $K_2$ (which have character 1 under $C_3$) for bands 3-4, $B_{C_3}(K -\mathbf{g}_1) = \mathbbm{1} = B_{C_3}(K'+\mathbf{g}_1-\mathbf{g}_2)$. This implies
\begin{equation}
C_3 W_3(1/2)C_3^{-1} = W_{K'\leftarrow (K-\mathbf{g}_1)} W_{K -\mathbf{g}_1\leftarrow K'} = \mathbbm{1},
\end{equation}
where we used the properties in Eq.~\eqref{eq:WLProperties}. Since $C_3$ is unitary, this implies $\{e^{i\theta_{3,3-4}(1/2)}\} = \{1,1\}$, and as a consequence, $\{e^{i\theta_{3-4}(1/2)}\} = \{1,1\}$. 

From our analysis above, we see that $\{e^{i\theta_{3,3-4}}\}$ go from $\{e^{i2\pi/3},e^{-i2\pi/3}\}$ to $\{1,1\}$ as $\Lambda$ goes from $0$ to $1/2$. Since the eigenvalues of $W_3(\Lambda)$ can only change continuously, the eigenvalues have to cross $e^{\pm i \pi/3}$ odd number of times. Hence, going from $\{1,1\}$ to $\{1,1\}$, the eigenvalues of $W_h(\Lambda)$ pass through $e^{\pm i \pi}$ odd number of times as $\Lambda$ goes from $0$ to $1/2$. This proves the winding seen in Fig.~\ref{fig:WL}(b) for bands 3-4.

In case of bands $1-2$, the story is the same at $\Lambda = 0$. However, at $\Lambda = 1/2$, $B_{C_3}(K -\mathbf{g}_1) \neq \mathbbm{1}$ since the the irrep label at the $K$ point for these two bands is $K_3$. Consequently, the eigenvalues of $W_3(1/2)$ can be any general phase $e^{\pm i \theta_3}$. This is why for bands $1-2$, the eigenvalues of $\log W_h$ are non-zero at $\Lambda = 1/2$ as can be seen in Fig.~\ref{fig:WL}(b). Similar things can be shown for bands $5-6$.

Our system corresponds to a triangular Bravais lattice, whose Wigner-Seitz unit cell along with its primitive lattice vectors is shown in Fig.~1(c). To numerically calculate the WL, we first descritize the BZ into $N$ three-fold rotation invariant loops and each loop into $n$ discrete points. We also parametrize the loops with $\Lambda$, distance of each loop from the HSP $\Gamma$, such that the smallest loop ($\Gamma$ point) and the biggest one (along the edges of the BZ) are respectively correspond to $\Lambda=0$ and $\Lambda=\sqrt{3}a\Lambda/2\pi$, where $a$ is the lattice constant, as illustrated in Fig.~\ref{fig:WL}(a). Then, for each loop at a fixed $\Lambda$, we calculate the eigenvectors at those $n$ points correspond to the desired sets of isolated modes. Let us denote the eigenvectors as $|u_i^{(j)} (\Lambda)\rangle$, where $i \in \{1, 2, …,n\}$ goes over the discrete points on the loop at a fixed $\Lambda$, and $j$ corresponds to the $j$th band under consideration; therefore, the WL expression for the two consecutive isolated $j$th and $j+1$th bands can then be written as follows:
 \begin{equation}\label{eq:wl}
 W(\Lambda)=\begin{bmatrix}
 \langle u_1^{(j)}|M|u_n^{(j)}\rangle & \langle u_1^{(j)}|M|u_n^{(j+1)}\rangle \\
 \langle u_1^{(j+1)}|M|u_n^{(j)}\rangle & \langle u_1^{(j+1)}|M|u_n^{(j+1)}\rangle 
 \end{bmatrix}.
  \prod_{i=1}^{n-1} \begin{bmatrix}
 \langle u_{i+1}^{(j)}|M|u_{i}^{(j)}\rangle & \langle u_{i+1}^{(j)}|M|u_{i}^{(j+1)}\rangle \\
 \langle u_{i+1}^{(j+1)}|M|u_{i}^{(j)}\rangle & \langle u_{i+1}^{(j+1)}|M|u_{i}^{(j+1)}\rangle 
 \end{bmatrix},
 \end{equation}
 where $W$ is a $2\times2$ matrix, and $M$ is the reduced mass matrix using for proper normalization. Let us consider $e_1(\Lambda), e_2(\Lambda)$ as two eigenvalues of the matrix W. Then, we calculate $\theta_1(\Lambda)=$ Im$(\ln(e_1))$ and $\theta_2(\Lambda)=$ Im$(\ln(e_2))$, where Im denotes the imaginary part. Finally, the evolution of $\theta(\Lambda)$ can be considered as symmetric WL. Likewise, Eqn.~\eqref{eq:wl} can be modified for any arbitrary $n$ numbers of isolated bands resulting in $n\times n$ $W$ matrix. We calculate symmetric WL for some sets of isolated bands of the Wigner-Seitz unit cell and the results are plotted in Fig.~\ref{fig:WL}(b). Fig.~\ref{fig:WL}(b) top right shows that the symmetric WL within the nontrivial bands (3-4) winds, while the addition of trivial bands (1-2) yield the non-winding WL depicted in Fig.~\ref{fig:WL}(b) bottom right, which verifies the fragility of these phases. We also calculate the non-winding WL within the trivial 1-2 and 5-6 isolated modes shown in Fig.~\ref{fig:WL}(b) left, top and bottom, respectively. 
\section{Irreps of trivial bands of wallpaper group $p31m$ at HSPS and detection of the fragility through simple band connectivity}\label{sec.3}
In the main text, we mentioned that if a set of isolated bands can be represented by Symmetric exponentially Localized Wannier Functions (SLWF) or atomic orbitals, the bands are trivial. Here, we show that no such isolated bands for wallpaper group $p31m$ have degeneracy at $\Gamma$ point and no degeneracy at $K$. To do that, we tabulate all such elementary band representations \cite{Bradlyn2017} obtained from SLWF in Tab.~\ref{tab:EBandRepp31m} (taken from BCS \cite{Bradlyn2017,vergniory2017graph,elcoro2017double}).

\begin{table}[h!]
    \caption{Elementary band representations and corresponding degeneracies at HSPs of wallpaper group $G = p31m$}
    \label{tab:EBandRepp31m}
    \centering
    \begin{tabular}{|c|c|c|}
        \hline
         type of orbital & irreps at HSPs & remarks \\
         \hline
         $s$-type orbital at Wyckoff pos. $1a$: $(A_1\uparrow G)_{1a}$ & $\Gamma_1-M_1-K_1$ & single band (no degeneracy at HSPs)\\
         \hline
         $f$-type orbital at Wyckoff pos. $1a$: $(A_2\uparrow G)_{1a}$ & $\Gamma_2-M_2-K_2$ & single band (no degeneracy at HSPs)\\
         \hline
         $(p_x,p_y)$-type orbitals at Wyckoff pos. $1a$: $(E\uparrow G)_{1a}$ & $\Gamma_3-M_1 \oplus M_2-K_3$ & double band (degeneracy at $\Gamma$ and $K$)\\
         \hline
         $s$-type orbital at Wyckoff pos. $2b$: $(A\uparrow G)_{2b}$ & $\Gamma_1\oplus\Gamma_2-M_1\oplus M_2-K_3$ & double band (degeneracy at $K$)\\
         \hline
         \multirow{3}{*}{$(p_x,p_y)$-type orbitals at Wyckoff pos. $2b$: $(^1E^2E\uparrow G)_{2b}$} & \multirow{3}{*}{$2\Gamma_3-2M_1\oplus 2M_2-K_1 \oplus K_2 \oplus K_3$} & $4$ bands (two doubly degenerate\\ & & eigenvalues at $\Gamma$ and one double\\
         & &  degeneracy at  $K$)\\
         \hline
    \end{tabular}
\end{table}

In Tab.~\ref{tab:EBandRepp31m}, $A_1$, $A_2$ and $E$ are the irreps of the site symmetry group $C_{3v}$ of Wyckoff position $1a$, and $A$ and $^1E^2E$ are the irreps of the site symmetry group $C_{3}$ of Wyckoff position $2b$ (see \cite{Bradlyn2017} for definition of Wyckoff positions and site symmetry group), and the orbitals $s$, $p_x$, $p_y$ and $f$ are just common names of the irreps. Clearly, there are no trivial set of two bands which has a degeneracy at the $\Gamma$ point, and no degeneracy at $K$. Consequently, the bands 3 and 4 in Fig.~1(d), having a degeneracy at the $\Gamma$ point and no other degeneracy at $K$, must be topological. On the other hand, there is only one fragile root for wallpaper group $p31m$ (see~\cite{song2020twisted} for details).
\begin{table}[h!]
    \caption{Fragile roots for wallpaper group $G = p31m$ with time reversal symmetry $T$ ($T^2 = 1$)}
    \label{tab:FragileRootp31m}
    \centering
    \begin{tabular}{|c|}
        \hline
         $\Gamma_3-M_1\oplus M_2-K_1\oplus K_2$\\
         \hline
    \end{tabular}
\end{table}

The irreps in Tab.~\ref{tab:FragileRootp31m} is the only way to have degeneracy which is allowed by compatibility relations between irreps (dictated by the mirror symmetry, see~\cite{dresselhaus2007group} for compatibility relations) at $\Gamma$ and no degeneracy at $K$ for a set of two bands. This means that a set of two isolated bands with a degeneracy at $\Gamma$ and no degeneracy at $K$ (with time reversal symmetry) is always fragile topological. This band connectivity of the lattice structure has been confirmed in our experiments as shown in Fig.~3(d).

\section{Explanation for the appearance of edge modes in the first and second BandGaps (BG)}\label{sec.4}
\begin{figure}[b]
    \centering
    \includegraphics[width = 0.7\textwidth]{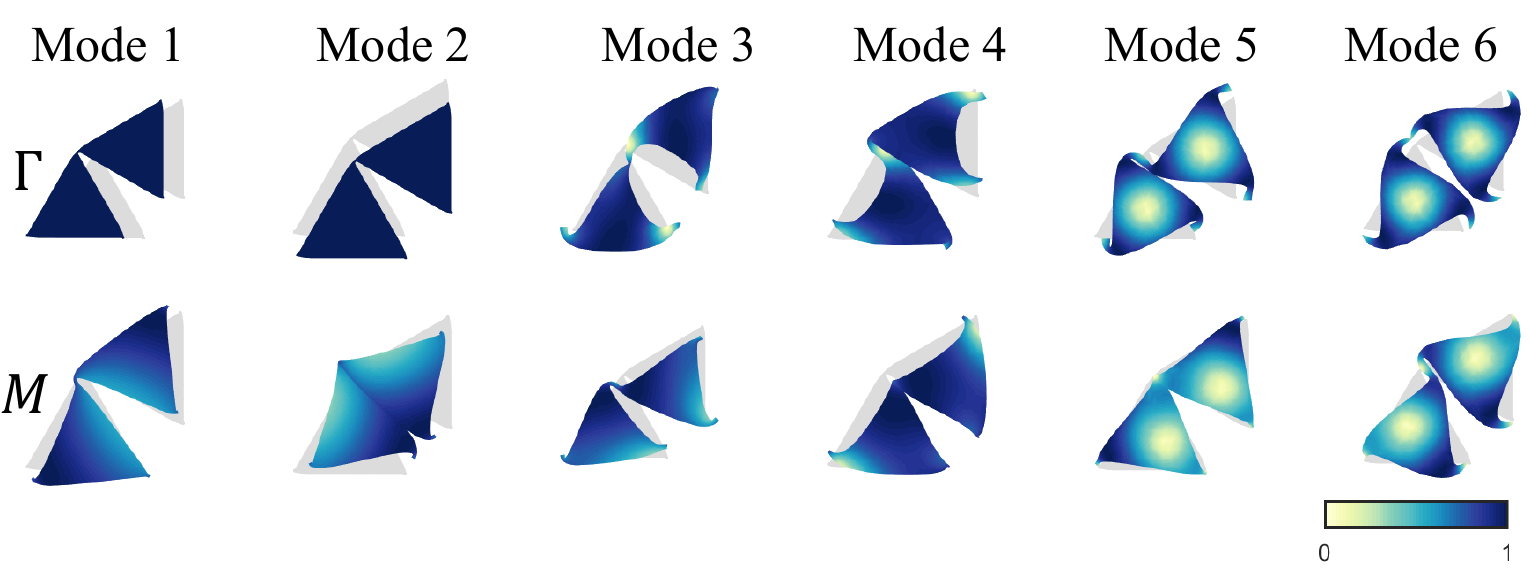}
    \caption{Deformation of the unit cell corresponding to the eigenfunctions at the HSPs $\Gamma$ (first row) and $M$ (second row) for the first six modes. Note that unlike the Wigner-Seitz unit cell in Fig.~\ref{fig:ModesWS}, this choice of unit cell is compatible with the supercell considered in Figs.~2(a-e) of the main text.}
    \label{fig:M,G}
\end{figure}
The origin of the edge modes in Figs.~2(b-e) can be understood in the following way. Starting from the band structure of the fully periodic system in Fig.~1(d) of the main text, if we create a super-cell with 15 unit cells in the $\mathbf{e}_1$ direction (still with fully periodic boundary condition), each band is going to fold 15 times. Since there are two bands in Fig.~1(d) below the first BG, there will be 30 bands below the first gap in the super-cell band structure. Similarly, between the first and second BG there will be 30 bands in the super-cell band structure. Now, if we open a boundary parallel to $\mathbf{e}_2$ of this super-cell, due to the open boundary condition, the band frequencies will change. Some bands may be pushed up or down in frequency from the bulk bands to go into the bulk gap and, as a result, will become edge bands. We see such two edge bands in the first BG in Fig.~2(a). We ask if those were pushed up from the bands below the gap or pushed down from the bands above. This can be answered easily by counting the number of bands below and above the first BG. We find that below BG1, there are 30 bands where between BG1 and BG2 there are 28 bands. This means that the two edge bands in BG1 were pushed down from bulk bands above BG1.

Another way to confirm this is to compare the mode shapes of the edge modes to the modes shapes of the bulk modes. We know that the $M$ point ($\Gamma$ and $M$ points) of the bulk single unit cell band structure gets projected to the $\xi = \pi$ ($\xi = 0$) of the super-cell band structure. Therefore, the edge modes at $\xi = \pi$ corresponding to the edge bands in BG1 should have the same character as (in other words, it should look like) the mode shape at the $M$ below or above the gap depending on if the edge gap is pushed up from the bulk below or pushed down from the bulk above, respectively. Comparison of the edge modes in Fig.~2(c) with the bulk modes in Fig.~\ref{fig:M,G} shows that the edge modes in Fig.~2(c) has the character of mode 3 at the $M$ point in Fig.~\ref{fig:M,G}. Similarly, a comparison of the edge modes in Fig.~2(b) ($\xi = 0$) with the bulk modes in Fig.~\ref{fig:M,G} shows that the edge modes in Fig.~2(b) has the character of mode 3 at the $M$ point in Fig.~\ref{fig:M,G}. These considerations confirm that the edge modes in the BG1 are actually pushed down from the bulk bands just above BG1. Similarly, there are only 28 bulk bands above BG 2 in Fig.~2(a) and the character of the edge modes in Figs.~2(d-e) is the same as that of mode 5 at $\Gamma$ and $M$ in Fig.~\ref{fig:M,G}, meaning that the edge modes in BG2 are pushed down from the bulk bands above BG2. This scenario is reminiscent of Rayleigh waves in elastic systems, which appear below the bulk acoustic bands.
\section{Material properties and specimen geometry}\label{sec.5}
\begin{figure}[b]
    \centering
    \includegraphics[width = 0.6\textwidth]{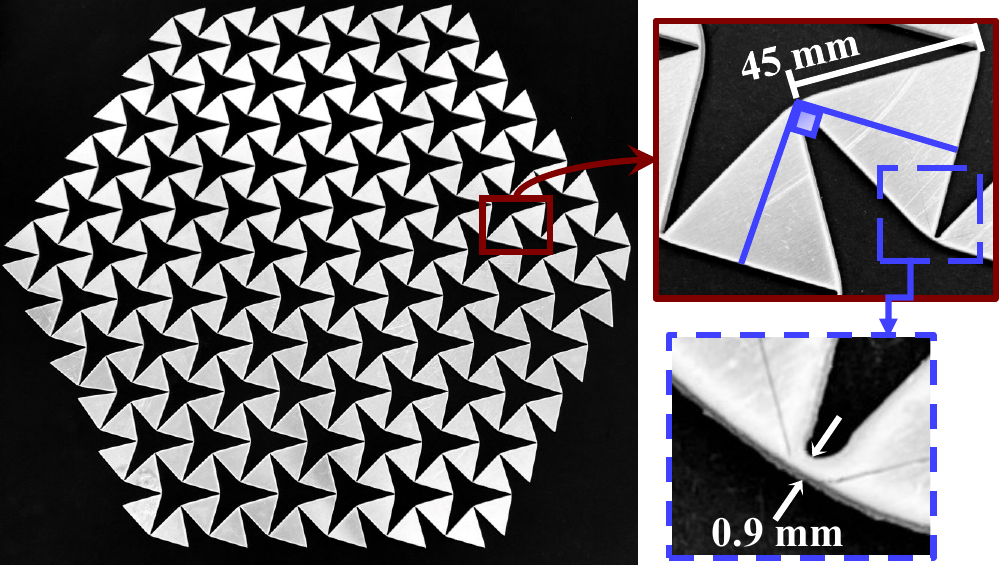}
    \caption{\label{fig:method} Hexagonal prototype manufactured via Water-jet cutting from Aluminum thin sheet, with insets indicating dimensions and showing details of the hinges.}
\end{figure}

We build a finite element model of the hexagonal metamaterial consisting of 108 cells, discretized with a mesh of plane-stress quadrilateral isoparametric elements, with appropriate mesh refinement in the neighborhood of the hinges where we expect concentration of deformation. This modeling is done using the software GMSH and exported as a “.STL” file for the manufacturing process. Fig.~\ref{fig:method} shows the geometry of the fabricated specimen obtained via water-jet cutting from a 2-mm thick sheet of Aluminum. The hexagonal domain has six unit cells on each edge. All the sides of the triangles in the kagome lattice and the widths of the ligaments constituting the structural hinges are 45 mm and 0.9 mm, respectively (see insets). The material properties of Aluminum are: Young’s modulus E = 71 GPa, Poisson’s ratio $\nu$ = 0.33 and density $\rho = 2700$ kg/m$^3$. 

\section{Experimental procedure for transmissibility analysis}\label{sec.6}
The setup depicted in Fig.~3(a) shows the  specimen vertically constrained via boundary supports and the three scanning laser heads of the (Polytec PSV 400 3D Scanning Laser Doppler Vibrometer (SLDV) required to acquire in-plane velocity measurements. The excitation is applied by an electromechanical shaker (Brüel \& Kj\ae r Type 4810) with an amplifier (Brüel \& Kj\ae r Type 2718) at the yellow dot probing the lattice surface. 
In-plane velocity is measured at three scan points per triangle. Retro-reflective tape is applied at the scan points to increase the reflectivity and reduce noise in the data. A broadband pseudo-random excitation is applied
to excite a broad range of frequencies. To supply enough energy at higher frequency to the shaker, we apply the signal in the 0-20 kHz range in two 10-kHz increments, prescribing higher amplitudes in the second step to yield roughly equal amount of energy injection for both intervals. 
\section{Band diagram reconstruction via 2D-DFT and wavefield acquisition}\label{sec.7}
For transient analysis, we modify the specimen setup such that it is vertically constrained only from one edge while the opposite edge is free as shown in the inset of Fig.~3(d). This setup allows better investigating the behavior of the corners, while still allowing measurements in the bulk. To reconstruct the band diagram, we separately apply 11 Hann-windowed burst excitations spanning both pass-band regions (with carrier frequencies of $\sim \, 2, 3, 4, 5, 8, 9, 10, 11, 12, 13, 13.7$ kHz) via a stinger probing the lattice in-plane at the bottom corner. We employ 5-cycle and 25-cycle bursts for the first 4 signals within the acoustic modes and the other 7 signals within the optical bands, respectively. The high number of cycle bursts help to excite narrower-band spectra, which is important in the reconstruction of the optical bands, where we seek evidence of fragile topology, since the branches are relatively flat in that range. We apply a high-pass filter (embedded in the vibrometer software) to filter out the low-frequency response dominated by spurious ambient vibrations. The in-plane and out-of-plane velocity components are measured by the laser heads at each of the 12 scan points 
denoted by red dots in Fig.~3(d), which lie along the $\Gamma-K$ direction. The time histories of the $\hat{x}$ and $\hat{y}$ in-plane components of the velocity, which correspond to lateral and axial components with respect to the selected direction, are then subjected to 2D-DFT, performed in MATLAB, (see \ref{s-6}) to obtain the spectral amplitude contour maps shown in Fig.~3(d). We use a similar setup to reconstruct the wavefield patterns established in the bulk and on the edges of the specimen. We use three different tone bursts with carrier frequencies of $\sim \, 3$, $\sim \, 6.85$, and $\sim \, 15$ kHz, lying in the pass-band, first BG and second BG, respectively. The measurements are performed at three scan points per each triangle of the lattice to strike a balance between the accuracy of the achieved spatial reconstruction of motion and the parsimony of the sensing strategy. As a standard procedure in time-domain laser vibrometry, we prescribe enough relaxation time to ensure that bursts have fully dissipated by damping before the next measurement is taken. The wavefield data is exported to MATLAB in the form of a $(\hat{x}-\hat{y}-\text{time})$ data cube. The slices of this cube represents wavefields at different time instants.
 
\section{Note on the discrepancy between the experiment and simulation results}\label{sec.8}
The observed mismatch between experiments and theory shown in Fig.~3(b), which is especially pronounced at the onset of the higher frequency gap, can be comfortably ascribable to the inevitable deviations of the specimen from the nominal characteristics of the model. Specifically, variability in the water-jet cutting process is responsible for a deviation/reduction of the average hinge thickness compared to the nominal 0.9 mm value, de facto softening the structure. Additionally, differences in material properties, geometric variabilities such as fluctuation of the hinges thickness from one cell to another and tapered thickness in the out-of-plane direction, and other non-idealities, including the formation of jagged edges during cutting process, also contribute to these discrepancies.

\section{2D Discrete Fourier Transform (2D-DFT)}\label{s-6}\label{sec.9}
\begin{figure}[h!]
    \centering
    \includegraphics[width = 0.7\textwidth]{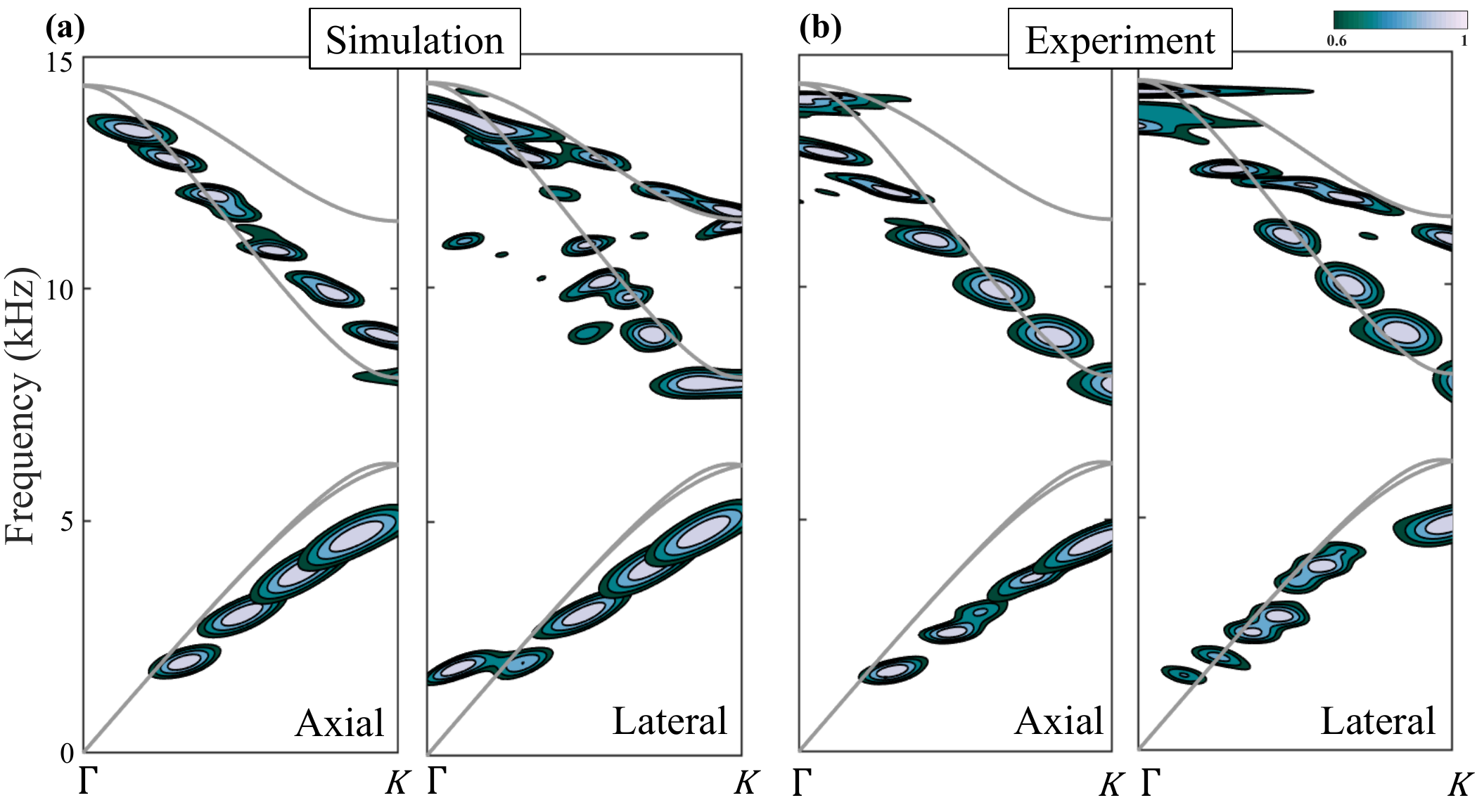}
    \caption{2D-DFT plots of the axial and lateral in-plane displacement (velocity) via simulation (a) (experiment (b)) carried out at several carrier frequencies}
    \label{fig:DFT}
\end{figure}

Figs.~\ref{fig:DFT}(a) and (b) show the reconstruction of the acoustic and optical branches of the band diagram from time histories over $n$ time instants collected at $m$ points located along the $\Gamma-K$ direction, from simulations and experimental data, respectively. 

The procedure involves 2D-DFT of the $m \times n$ spatio-temporal data matrix encompassing the time histories at all the selected points stacked as columns. The outcome of the 2D-DFT is a matrix of spectral amplitudes, which represents the dependence of spectral amplitudes upon frequency and a scalar wave number sampled along $\Gamma-K$. The simulation data consists of nodal displacements from the FEM transient simulations, while the experimental data are velocities measured by the vibrometer at the available scan points. For each point, we have two components of the displacement(velocity) vector, an \textit{axial} component parallel to the $\Gamma-K$ direction and a \textit{lateral} one perpendicular to it. The spectral amplitude contours obtained form the different bursts are superimposed to the band diagram obtained via unit cell analysis to highlight the most relevant mode activation at each frequency and obtain a piece-wise reconstruction of the band diagram branches.

Overall, the qualitative  behavior of the reconstructed acoustic modes, including the folding at the edge of the BZ, is captured correctly. Quantitatively, we report a frequency discrepancy that becomes more pronounced as we approach the onset of the BG. This discrepancy may be in part attributed to spurious boundary effects that are not accounted for in the band diagram (which assumes an infinite lattice) but can be conspicuous in the finite-domain transient simulations, especially working with a lattice with low unit cell count. The frequency shifting can also be in part associated with the fluctuations of material properties that have already been discussed as a probable cause of the downward shift of the BG onset in the steady-state transmissibility analysis. Interestingly, the agreement is far superior for the optical branches, which are of interests for the discussion on fragile topology reported in the main Letter. A complete understanding of the reasons for the different degree of agreement between acoustic and optical modes is still missing and warrants additional investigation in future studies.
In general, it can be seen that the lateral component of the displacement(velocity) vector yields a better agreement, which suggests that shear-like mechanisms may be dominant in the modal characters of these branches.

\end{document}